\documentclass[journal=jpccck,manuscript=article]{achemso}

\usepackage{chemformula} 
\usepackage[separate-uncertainty=true,separate-uncertainty,multi-part-units=single]{siunitx} 
\usepackage{array}
\usepackage{multirow}
\usepackage[colorlinks=true,linkcolor=blue,urlcolor=blue, citecolor=blue]{hyperref}
\usepackage{chemformula} 
\usepackage{stackrel,amssymb,amsmath}

\author{Haibo Xue}
    \affiliation{Materials Simulation \& Modelling, Department of Applied Physics, Eindhoven University of Technology, P.O. Box 513, 5600MB Eindhoven, the Netherlands.}
    \alsoaffiliation{Center for Computational Energy Research, Department of Applied Physics, Eindhoven University of Technology, P.O. Box 513, 5600MB Eindhoven, the Netherlands.}

\author{Jos\'{e} Manuel Vicent-Luna}
    \affiliation{Materials Simulation \& Modelling, Department of Applied Physics, Eindhoven University of Technology, P.O. Box 513, 5600MB Eindhoven, the Netherlands.}
    \alsoaffiliation{Center for Computational Energy Research, Department of Applied Physics, Eindhoven University of Technology, P.O. Box 513, 5600MB Eindhoven, the Netherlands.}
    
\author{Shuxia Tao}
    \affiliation{Materials Simulation \& Modelling, Department of Applied Physics, Eindhoven University of Technology, P.O. Box 513, 5600MB Eindhoven, the Netherlands.}
    \alsoaffiliation{Center for Computational Energy Research, Department of Applied Physics, Eindhoven University of Technology, P.O. Box 513, 5600MB Eindhoven, the Netherlands.}
    \email{s.x.tao@tue.nl}
    
\author{Geert Brocks}
    \affiliation{Materials Simulation \& Modelling, Department of Applied Physics, Eindhoven University of Technology, P.O. Box 513, 5600MB Eindhoven, the Netherlands.}
    \alsoaffiliation{Center for Computational Energy Research, Department of Applied Physics, Eindhoven University of Technology, P.O. Box 513, 5600MB Eindhoven, the Netherlands.}
    \alsoaffiliation{Computational Materials Science, Faculty of Science and Technology and MESA+ Institute for Nanotechnology, University of Twente, P.O. Box 217, 7500AE Enschede, the Netherlands.}
    \email{g.h.l.a.brocks@utwente.nl}

\title{Compound Defects in Halide Perovskites:\\ A First-Principles Study of CsPbI$_3$}


\begin{document}

\begin{tocentry}

\includegraphics{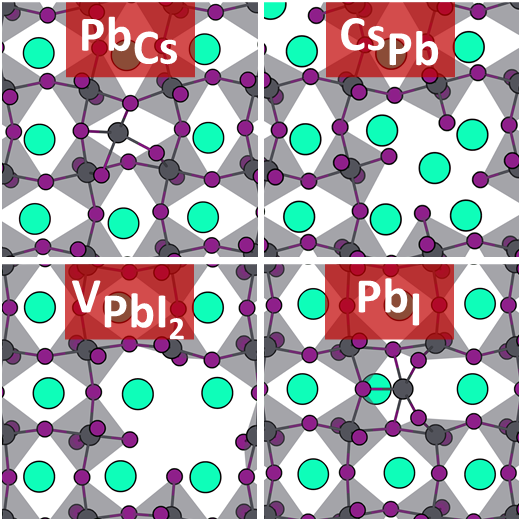}

\end{tocentry}

\begin{abstract}

Lattice defects affect the long-term stability of halide perovskite solar cells. Whereas simple point defects, i.e., atomic interstitials and vacancies, have been studied in great detail, here we focus on compound defects that are more likely to form under crystal growth conditions, such as compound vacancies or interstitials, and antisites. We identify the most prominent defects in the archetype inorganic perovskite CsPbI$_3$, through first-principles density functional theory (DFT) calculations. We find that under equilibrium conditions at room temperature, the antisite of Pb substituting Cs forms in a concentration comparable to those of the most prominent point defects, whereas the other compound defects are negligible. However, under nonequilibrium thermal and operating conditions, other complexes also become as important as the point defects. Those are the Cs substituting Pb antisite, and, to a lesser extent, the compound vacancies of PbI$_2$ or CsPbI$_3$ units, and the I substituting Cs antisite. These compound defects only lead to shallow or inactive charge carrier traps, which testifies to the electronic stability of the halide perovskites. Under operating conditions with a quasi Fermi level very close to the valence band, deeper traps can develop.  

\end{abstract}


\section{Introduction}

On the basis of their outstanding efficiency (25.7\% to date \cite{NREL2022}) and relative ease of fabrication, halide perovskite solar cells seem to be poised for large scale applications. The primary obstacle blocking their present commercialization is their relative rapid degradation under operating conditions \cite{Kim2016, Zhouyuanyuan2019, Park2019, He2020}. On a microscopic level, lattice defects in the perovskite materials initiate the degradation process, as they facilitate migration of ions \cite{Yuan2016, Yang2016a, Eames2015, Azpiroz2015a, Pols2021}, chemical reactions \cite{Mosconi2015, Aristidou2017}, phase transitions \cite{Tan2020}, and phase segregation \cite{Barker2017}.

Because of the experimental difficulties in characterizing defect structures microscopically, much of our current understanding of lattice defects in halide perovskites stems from results obtained from electronic structure calculations based on density functional theory (DFT). Following  common semiconductor practice \cite{Walle2004, Freysoldt2014}, elementary defects consisting of single atomic interstitials, vacancies, or antisites have been at the center of interest \cite{Yin2014, Shi2015a, Meggiolaro2018review, Huang2018, Meggiolaro2019}. In a previous work \cite{Xue2022perovskites}, we have studied vacancy and interstitial point defects in six primary Pb- and Sn-based halide perovskites with different cations (Cs, MA, FA) and anions (I, Br, Cl), within a single computational framework \cite{Xue2021functionals}. One prevalent conclusion from most of these computational studies is that in these materials the points defects with the highest concentrations under equilibrium growth conditions, only introduce shallow traps. 

Defects in halide perovskites with a more complex structure have also been considered \cite{Kye2018, Kim2014, Keeble2021}. Conceptually, such complex defects can be thought of as resulting from a recombination of simple atomic point defects (vacancies or interstitials) to, for instance, PbI$_2$ or MAI compound vacancies in MAPbI$_3$ \cite{Kim2014, Kye2018, Keeble2021}. Within this line of thought also antisites can be interpreted as compound defects, resulting from a recombination of an interstitial and a vacancy of different species. For instance, in CsPbI$_3$, cation antisites result from a recombination of a Cs vacancy (interstitial) and an Pb interstitial (vacancy) \cite{Huang2018}. Some compound defects have been predicted to form shallow defects only \cite{Kim2014, Huang2018}, whereas others have the potential to form deep traps \cite{Kye2018, Michael2014}.

The formation energy of compound defects is typically much higher than that of simple point defects, which implies that under normal equilibrium conditions (room temperature, atmospheric pressure) the concentration of compound defects, including antisites, is negligible  \cite{Walle2004, Freysoldt2014, Yin2014, Shi2015a}. However, many crystal growth conditions are highly nonequilibrium, and defects can be formed during growth in appreciable concentrations \cite{Xue2022perovskites}. In molecular dynamics simulations that use a reactive force field \cite{Pols2021, Pols2022}, applied to halide perovskites containing an appreciable amount of point defects, one often observes recombination of the latter to compound defects. From positron annihilation lifetime spectroscopy, assisted by DFT calculations, there is evidence of charge carrier trapping at compound vacancy defects in MAPbI$_3$ \cite{Keeble2021}.  

In this paper, we study compound defects, vacancies, interstitials and antisites, in the archetype inorganic perovskite CsPbI$_3$ by means of first-principles DFT calculations. Not only do we calculate their equilibrium concentrations, but explicitly considering the possible recombination reactions of elementary point defects, we also assess their concentrations under nonequilibrium conditions. The effect of these compound defects on the electronic properties is examined, in particular their potential to form deep traps.

\section{Computational Methods}

\subsection{DFT calculations}

Density functional theory (DFT) calculations are performed with the Vienna \textit{Ab-Initio} Simulation Package (VASP) \cite{Kresse1993, Kresse1996, Kresse1996a}, employing the SCAN+rVV10 \cite{Peng2016, Xue2021functionals} functional for electronic calculations and geometry optimization. Our calculations use a plane wave kinetic energy cutoff of 500 eV and a $\Gamma$-point-only \textbf{k}-point mesh. The energy and force convergence criteria are set to 10$^{-4}$ eV and 0.02 eV/\AA, respectively. Spin-orbit coupling is omitted, as it has little effect on the formation energies of defects \cite{Xue2021functionals}. Spin polarization is included in all calculations.

As in our previous work \cite{Xue2022perovskites}, point defects or compound defects are created in a $2 \times 2 \times 2$ orthorhombic supercell of CsPbI$_3$, which contains 32 formula units. The lattice volume and ionic positions of the pristine supercell are fully relaxed. Within the supercell, atomic positions of defective structures are optimized. For antisite defects, we use the notation $A_{B}$ to indicate that atom $A$ substitutes atom $B$ in the lattice. We consider twelve different compound complexes, i.e., the three compound vacancies $\mathrm{V_{[CsI]}}$, $\mathrm{V_{[PbI_2]}}$, and $\mathrm{V_{[CsPbI_3]}}$, the two compound interstitials $\mathrm{[CsI]_i}$ and $\mathrm{[PbI_2]_i}$, the six antisites $\mathrm{Cs_I}$, $\mathrm{I_{Cs}}$, $\mathrm{Cs_{Pb}}$, $\mathrm{Pb_{Cs}}$, $\mathrm{Pb_I}$, and $\mathrm{I_{Pb}}$, and the compound antisite $\mathrm{[2Cs]_{Pb}}$. 

\subsection{Defect Formation Energy}

Under equilibrium conditions, the concentrations of lattice defects can be obtained from Boltzmann statistics
\begin{equation} \label{eq: defect concentration}
    \frac{c(D^q)}{c_0(D^q)-c(D^q)} =  \exp \left[- \frac{\Delta H_f (D^q)}{k_BT} \right],
\end{equation}
where $D^q$ indicates the type of defect, either a simple interstitial or vacancy point defect, or a compound interstitial or vacancy, or an antisite defect, with charge $q$; $c$ is the defect concentration, and $c_0$ is the density of possible sites for that particular defect (including orientational degrees of freedom if the defect in not spherically symmetric), where usually $c \ll c_0$; $\Delta H_f$ is the defect formation energy (DFE), $T$ is the temperature, and $k_B$ is the Boltzmann constant. 

Different types of defects have different charges, but if no external charges are injected, then as a whole a material  has to be charge neutral
\begin{equation} \label{eq:charge_neutrality}
    p - n + \sum_{D^q} q \; c(D^q) = 0,
\end{equation}
where $p$ and $n$ are the intrinsic charge densities of holes and electrons of the semiconductor material. The charge neutrality condition, Eq. (\ref{eq:charge_neutrality}), fixes the intrinsic Fermi level. 

The DFE is calculated from the expression \cite{Walle2004}
\begin{equation}\label{eq: DFE}
    \begin{aligned}
        \Delta H_{f}(D^q) = &E_\mathrm{tot}(D^q)-E_\mathrm{bulk} - \sum_{k} n_{k} \  \mu_{k} \\ 
        &+ q(E_{F}+ E_\mathrm{VBM}+ \Delta V),
    \end{aligned}
\end{equation}
where $E_\mathrm{tot}$ and $E_\mathrm{bulk}$ are the DFT total energies of the defective and pristine supercells, respectively, and $n_k$ and $\mu_k$ are the number of atoms and chemical potential of atomic species $k$ added to ($n_k > 0$) or removed from ($n_k<0$) the pristine supercell in order to create the defect. We use the chemical potentials $\mu_k$; $k=$ Cs,Pb,I, as determined for I-medium conditions in our previous work \cite{Xue2022perovskites}. 

Creating a charge $q$ requires taking electrons from or adding them to a reservoir at a fixed Fermi level. The latter  is calculated as $E_{F}+ E_\mathrm{VBM}$, with $0\leq E_F \leq E_g$, the band gap, and $E_\mathrm{VBM}$ the energy of the valence band maximum. As it is difficult to determine the latter from a calculation on a defective cell, one establishes $E_\mathrm{VBM}$ in the pristine cell, shifted by $\Delta V$, which is calculated by lining up the core level on an atom in the pristine and the neutral defective cell that is far from the defect \cite{Walle2004, Komsa2012}. The supercell in the calculation and the dielectric screening in CsPbI$_3$ are sufficiently large, so that the electrostatic interaction between a charged point defect and its periodically repeated images can be neglected \cite{Meggiolaro2018review, Xue2021functionals,Xue2022perovskites}. In addition, we neglect vibrational contributions to the DFEs, and the effect of thermal expansion on the DFEs, as these are typically small in the present compounds \cite{Xue2021functionals, Wiktor2017}.

\subsection{Recombination Reaction}
We model the recombination of point defects $A_1,...,A_m$ to a compound defect $B$ as a chemical reaction
\begin{equation}\label{eq: chemical reaction}
    \alpha_1 A_1 +  \alpha_2 A_2 + ... +  \alpha_{m} A_{m} \leftrightarrows \beta B.   
\end{equation}
Reaction equilibrium is defined by
\begin{equation}\label{eq: equilibrium}
    \sum_{i=1}^m \alpha_i \mu_i = \beta \mu_B,   
\end{equation}
with $\mu_i$ and $\mu_B$ the chemical potentials of species $A_i$ and $B$, given by 
\begin{align}
   &\mu_i =  \Delta H_{f}(A_i^{q_i}) +  k_B T \ln \frac{c_i}{c_{0,i}-c_i}; \ \ \ i=1,...,m; \label{eq: chemical potential1} \\
   &\mu_B =  \Delta H_{f}(B^{q_B}) +  k_B T \ln \frac{c_B}{c_{0,B}-c_B}, \label{eq: chemical potential2}
\end{align}
where $\Delta H_{f}(D^q)$; $D^q=A_i^{q_i},B^{q_B}$ are the DFEs according to Eq. (\ref{eq: DFE}), $c_i = c(A_i^{q_i})$, $c_B = c(B^{q_B})$ are concentrations, and $c_{0,i} = c_0(A_i^{q_i})$, $c_{0,B} = c_0(B^{q_B})$ are the densities of possible sites (see Table S1 of the Supporting Information for details). Note that we do not assume that charge is conserved in reaction (\ref{eq: chemical reaction}). The electron reservoir with Fermi energy $E_F$ can supply electrons or holes, which is accounted for in the DFEs. Equations (\ref{eq: equilibrium})-(\ref{eq: chemical potential2}) give the law of mass action
\begin{equation}
   \left(\frac{c_B}{c_{0,B}-c_B}\right)^\beta \prod_{i=1}^{m} \left(\frac{c_{0,i}-c_i}{c_i}\right)^{\alpha_i} = \exp\left[ -\frac{\Delta H_r}{k_B T}\right], \label{eq: mass action} 
\end{equation}
\begin{equation}
   \Delta H_r = \beta \Delta H_{f}(B^{q_B}) - \sum_{i=1}^{m} \alpha_i \Delta H_{f}(A_i^{q_i}), \label{eq: delta H}
\end{equation}
where $\Delta H_r$ is the reaction energy of reaction (\ref{eq: chemical reaction}).

If all (simple and compound) defects are in equilibrium with reservoirs at chemical potentials $\mu_k$, Eq. (\ref{eq: DFE}), then their concentrations are given by Eq. (\ref{eq: defect concentration}), and trivially obey the law of mass action, Eq. (\ref{eq: mass action}). Typically, however, point defects and compound defects are initially created at concentrations $c_i^{(0)}$ and $c_B^{(0)}$, respectively, in a crystal growth process, after which the crystals are extracted and kept at room temperature. The defects then remain, but they can recombine according to Eq. (\ref{eq: chemical reaction}). Not only does this include the possible formation of compound interstitials or vacancies, but also the formation of antisites through the recombination of an interstitial and a vacancy.  

As the recombination reaction, Eq. (\ref{eq: chemical reaction}), conserves the total number of atoms of each species, one has 
\begin{equation}\label{eq:conservation}
    c_i + \frac{\alpha_i}{\beta} c_B = c_i^{(0)} + \frac{\alpha_i}{\beta} c_B^{(0)};\ \ \ i=1,...,m.   
\end{equation}
Given the initial concentrations $c_i^{(0)}$ and $c_B^{(0)}$, the law of mass action, Eq. (\ref{eq: mass action}), then allows for determining the actual concentrations of the compound defect $c_B$, and of the point defects $c_i$. 

\subsection{Charge State Transition Level}
Under operating conditions, charges are injected in the material, shifting the positions of the (quasi) Fermi levels for electrons and holes. The charge state transition level (CSTL) $\varepsilon(q/q')$ is defined as the Fermi level position where the charge states $q$ and $q'$ of the same type of defect have equal formation energy, $\Delta H_{f}(D^q)=\Delta H_{f}(D^{q'})$. As the DFEs have a simple linear dependence on $E_F$, Eq. (\ref{eq: DFE}), this condition can be expressed as
\begin{equation} \label{eq: CSTL}
    \varepsilon(q/q')= \frac{\Delta H_{f}(D^q,E_{F}=0)-\Delta H_{f}(D^{q'},E_{F}=0)}{q'-q},
\end{equation}
where $\Delta H_{f}(D^q,E_{F}=0)$ is the DFE calculated at $E_F = 0$. The CSTLs are important for the electronic properties; if these levels are deep inside the band gap, they can trap charge carriers, and act as nonradiative recombination centers.

The band gap calculated with SCAN+rVV10 suffers from the DFT band gap error. However, we would argue that the positions of the CSTLs with respect to the band edges are correct, because the defects' electronic states have a character similar to either the valence band or the conduction band \cite{Tao2019}. For a more detailed discussion, see Ref. \citenum{Xue2022perovskites}.

\section{Results and discussion}

\subsection{Equilibrium Conditions}
\begin{figure}
    \includegraphics[width=15cm]{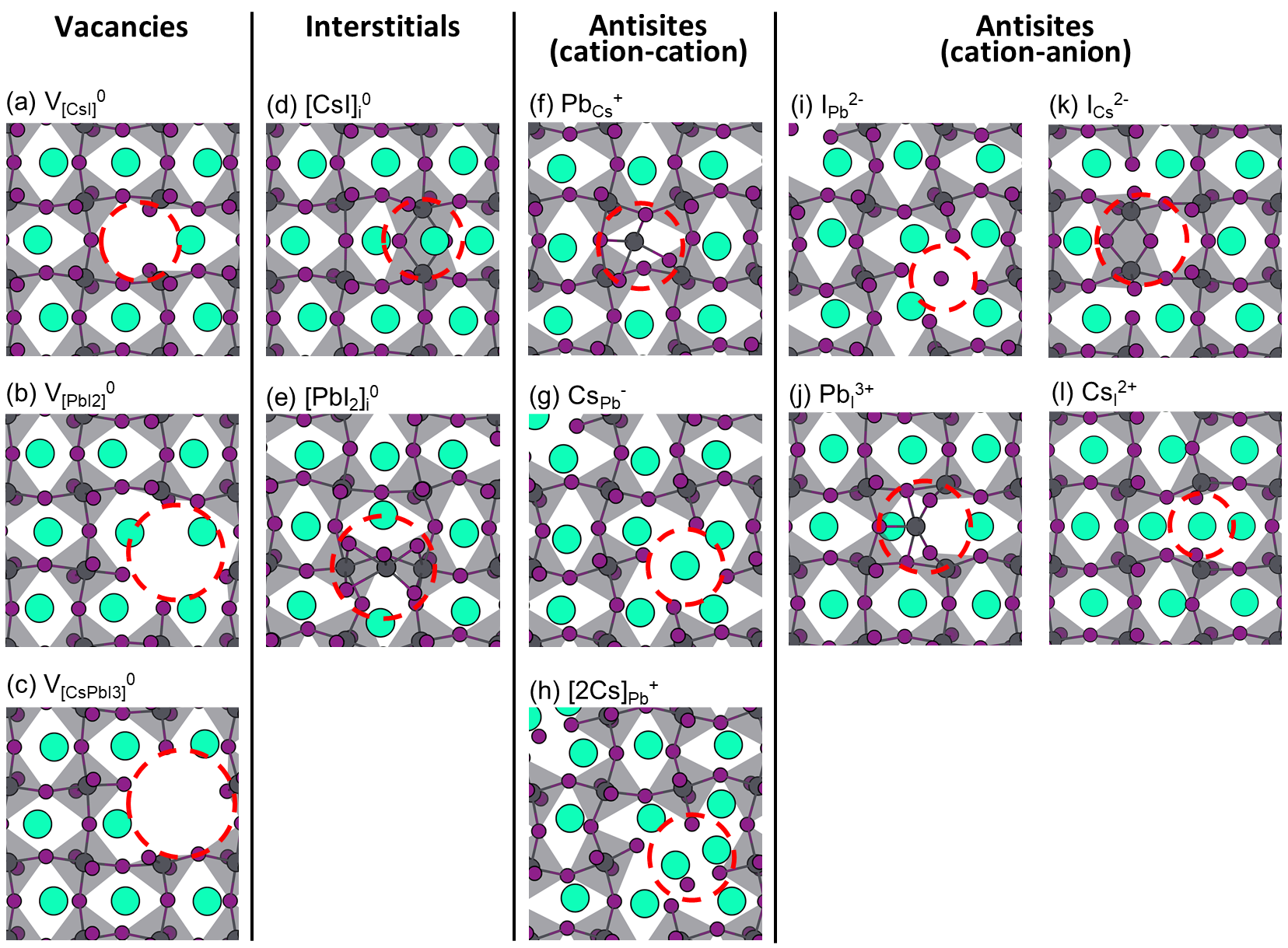}
    \caption{Optimized structures of compound defects in CsPbI$_3$ in their most stable charge states under intrinsic conditions; (a-c) compound vacancies, (d-e) compound interstitials, (f-h) cation-cation antisites, and (i-l) cation-anion antisites. Cs, Pb and I atoms are represented by green, black and purple circles, respectively, with PbI octahedra colored gray. The positions of the compound defects are indicated by the red circles.} 
    \label{fig: structures}
\end{figure}

Formation of compound defects in semiconductors is often driven by the attractive electrostatic interaction between defects with opposite charge states \cite{Walle2004, Freysoldt2014}.  Possible compound vacancies in CsPbI$_3$, resulting from recombination of the point vacancies $\mathrm{V_{Cs}}^{-}$, $\mathrm{V_{Pb}}^{2-}$ and $\mathrm{V_{I}}^{+}$,  are then $\mathrm{V_{CsI}}$, $\mathrm{V_{PbI_2}}$ and $\mathrm{V_{[CsPbI_3]}}$, where the neutral state indeed turns out to be the most stable charge state under intrinsic conditions. Optimized structures of these defects are shown in Figures \ref{fig: structures}(a-c). 

Following the same reasoning, we find the neutral compound interstitial defects $\mathrm{[CsI]_i}$ and $\mathrm{[PbI_2]_i}$, shown in Figures \ref{fig: structures}(d,e) through recombination of the point interstitials $\mathrm{Cs_i}^+$, $\mathrm{Pb_i}^{2+}$, and $\mathrm{I_i}^-$. For larger potential compound interstitials, such as $\mathrm{[CsPbI_3]_i}$, we found that the lattice becomes too distorted and the DFE becomes very large.

\begin{figure}
    \includegraphics[width=14cm]{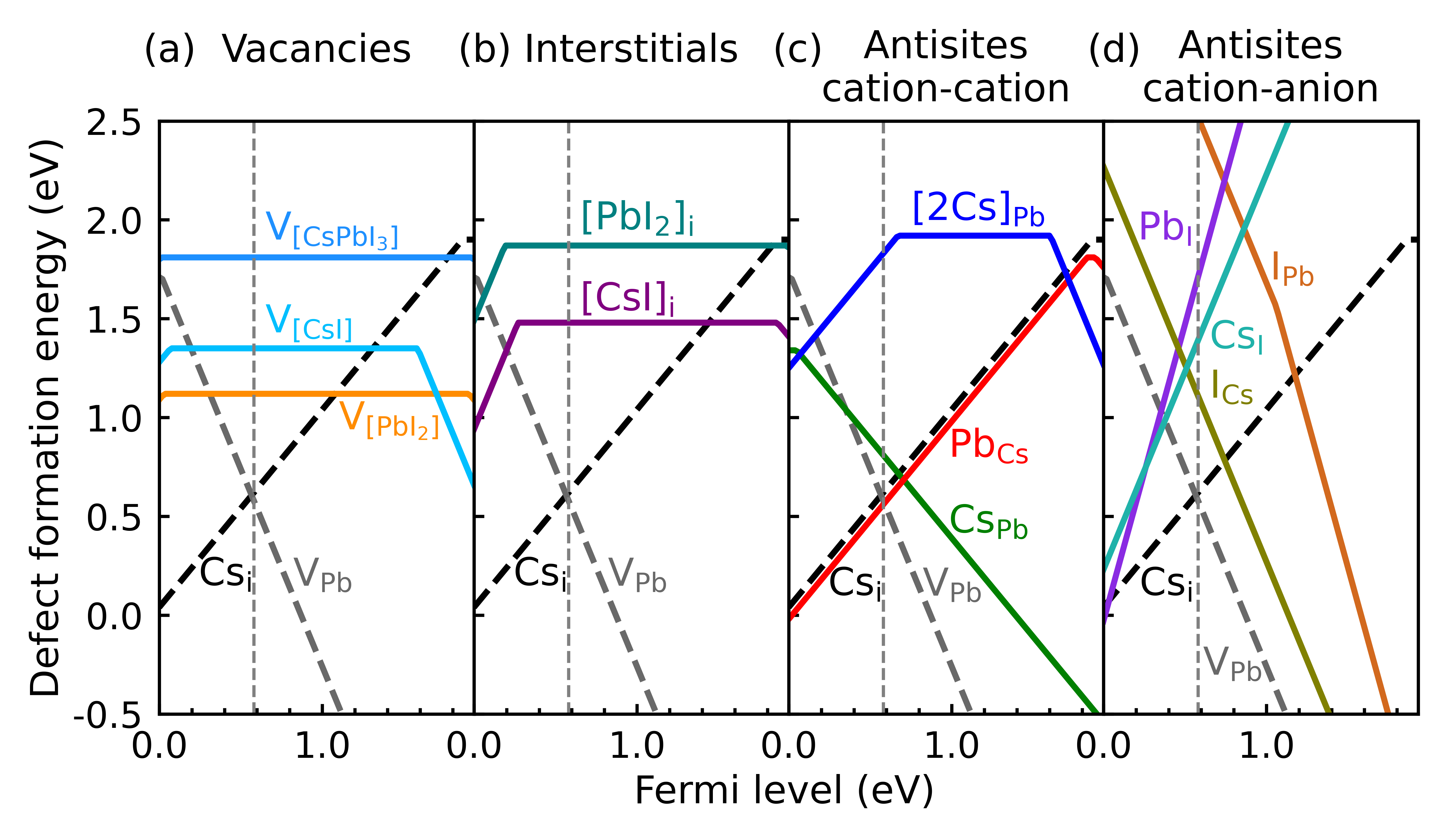}
    \caption{Formation energies of compound defects as a function of the Fermi level; (a) vacancies, (b) interstitials, (c) cation-cation antisites, and (d) cation-anion antisites. For comparison, the dashed black lines represent the formation energies of the two dominant point defects in CsPbI$_3$. The intrinsic Fermi level, determined by the charge neutrality condition, Eq. (\ref{eq:charge_neutrality}), is indicated by the vertical dashed gray line.}
    \label{fig: DFEs}
\end{figure}

Formation energies of the compound vacancies and interstitials, calculated according to Eq. (\ref{eq: DFE}), are shown in Figures \ref{fig: DFEs}(a,b). Taking into account of all point defects and compound defects, the intrinsic Fermi level, $E_F^{(i)}$, calculated with the charge neutrality condition, Eq. (\ref{eq:charge_neutrality}), is 0.58 eV with respect to the VBM. At this condition $\mathrm{Cs_i}^+$ and $\mathrm{V_{Pb}}^{2-}$ are the dominant atomic point defects \cite{Xue2022perovskites}, and the antisite $\mathrm{Pb_{Cs}}^+$, to be discussed below, is the most dominant compound defect. The compound vacancy and interstitial defects listed above, are then all stable in the neutral state. A list of the DFEs and concentrations, calculated at the intrinsic Fermi level, of these compound defects is given in Table \ref{table DFEs}. 

A compound vacancy defect creates a considerable hole in the lattice, see Figures \ref{fig: structures}(a-c), and its DFE is correspondingly high. The vacancy $\mathrm{V_{PbI_2}}$ is relatively easiest to form, with a DFE of 1.12 eV, followed by $\mathrm{V_{CsI}}$ and $\mathrm{V_{[CsPbI_3]}}$, whose DFEs are 1.35 eV and 1.81 eV, respectively. All of these numbers are $\gtrsim 0.5$ eV higher than the DFEs of the simple point defects $\mathrm{Cs_i}^+$ and $\mathrm{V_{Pb}}^{2-}$, which means that concentration of compound vacancy defects is negligible at room temperature under equilibrium conditions (Table \ref{table DFEs}).

Compound interstitial defects, $\mathrm{[CsI]_i}$ and $\mathrm{[PbI_2]_i}$, can be accommodated in the CsPbI$_3$ lattice by a distortion or tilting of the Pb-I octahedra, see Figures \ref{fig: structures}(d,e), albeit at a considerable energy penalty, with DFEs of 1.48 eV and 1.87 eV, respectively. We conclude that compound interstitial defects also have negligible concentrations at room temperature under equilibrium conditions, see Table \ref{table DFEs}.

Turning to antisite defects, as there are two different cations in CsPbI$_3$, antisites can be formed among cations, i.e., by a cation of one type occupying a position of a cation of the other type, $\mathrm{Pb_{Cs}}$ (Pb substitutes Cs), or $\mathrm{Cs_{Pb}}$, see Figures \ref{fig: structures}(f, g). We stick to the nomenclature of antisites, but note that replacing one cation by another can lead to a notable local distortion of the lattice, such that the substituting ion is significantly displaced from the lattice position of the original ion. Pb ions are nominally 2+, and Cs ions are 1+, so it is not surprising to find the most stable charge states of these antisites as $\mathrm{Pb_{Cs}}^+$ and $\mathrm{Cs_{Pb}}^-$. The DFE of $\mathrm{Pb_{Cs}}^+$ is comparable to that of the simple point defects $\mathrm{Cs_i}^+$ and $\mathrm{V_{Pb}}^{2-}$, see Figure \ref{fig: DFEs}(c) and Table \ref{table DFEs}, which means that this antisite occurs relatively frequently under equilibrium conditions. The $\mathrm{Cs_{Pb}}^-$ antisite defect has a DFE that is $\sim$ 0.25 eV larger than that of $\mathrm{Pb_{Cs}}^+$, making it less favorable. 

\begin{table} 
    \caption{Formation energies ($\Delta H_f$) and concentrations of compound defects under equilibrium conditions ($c_{\mathrm{equilibrium}}$, $T=300$K); recombination reactions, reaction energies ($\Delta H_r$) and concentrations of compound defects under nonequilibrium conditions  \textsuperscript{\emph{a}} ($c_{\mathrm{nonequilibrium}}$), all at the intrinsic Fermi level.}
    \label{table DFEs}
    \begin{tabular}{llllll}
        \hline
        \multirow{2}{*}{Defects} & $\Delta H_f$ & $c_{\mathrm{equilibrium}}$ & \multirow{2}{*}{Reaction} & $\Delta H_r$ & $c_{\mathrm{nonequilibrium}}$\\
         & (eV) & (cm$^{-3}$) & & (eV) & (cm$^{-3}$) \\
        \hline
        \multicolumn{4}{l}{\underline{Vacancies}} \\
        $\mathrm{V_{CsI}}^0$ & 1.35 & $2.81 \times 10^{-1}$ & $\mathrm{V_{Cs}}^- + \mathrm{V_{I}}^+$ &$-0.06$ & $7.75 \times 10^{8}$ \\
        $\mathrm{V_{PbI_2}}^0$ & 1.12 & $7.23 \times 10^{2}$ & $\mathrm{V_{Pb}}^{2-} + 2\mathrm{V_{I}}^+$ &$-0.83$ & $5.26 \times 10^{14}$ \\
        $\mathrm{V_{CsPbI_3}}^0$ & 1.81 & $1.73 \times 10^{-9}$ & $\mathrm{V_{Cs}}^- + \mathrm{V_{Pb}}^{2-} + 3\mathrm{V_{I}}^+$ &$-1.54$ & $5.10 \times 10^{13}$\\
        \multicolumn{4}{l}{\underline{Interstitials}} \\
        $\mathrm{[CsI]_i}^0$ & 1.48 & $1.51 \times 10^{-3}$ & $\mathrm{Cs_{i}}^+ + \mathrm{I_i}^-$ & 0.16 & $1.41 \times 10^{6}$\\
        $\mathrm{[PbI_2]_i}^0$ & 1.87 & $1.09 \times 10^{-9}$ & $\mathrm{Pb_{i}}^{2+} + 2\mathrm{I_i}^-$ & $-0.30$ & $3.24 \times 10^{5}$\\
        \multicolumn{4}{l}{\underline{Antisites (cation-cation)}} \\
        $\mathrm{Pb_{Cs}}^+$ & 0.56 & $1.71 \times 10^{12}$ & $\mathrm{Pb_{i}}^{2+} + \mathrm{V_{Cs}}^-$ & $-0.93$ & $9.44 \times 10^{15}$\\
        $\mathrm{Cs_{Pb}}^-$ & 0.81 & $1.09 \times 10^{8}$ & $\mathrm{Cs_{i}}^+ + \mathrm{V_{Pb}}^{2-}$ & $-0.39$ & $3.88 \times 10^{15}$\\
        $\mathrm{[2Cs]_{Pb}}^+$ & 1.83 & $7.72 \times 10^{-10}$ & $2\mathrm{Cs_{i}}^+ + \mathrm{V_{Pb}}^{2-} + e^+$ & 0.01 & $1.51 \times 10^{3}$\\
        \multicolumn{4}{l}{\underline{Antisites (cation-anion)}} \\
        $\mathrm{I_{Pb}}^{2-}$ & 2.51 & $2.90 \times 10^{-21}$ & $\mathrm{I_i}^- + \mathrm{V_{Pb}}^{2-} + e^+$ & 1.24 & $1.12 \times 10^{-12}$\\ 
        $\mathrm{Pb_I}^{3+}$ & 1.72 & $1.58 \times 10^{-7}$ & $\mathrm{Pb_{i}}^{2+} +  \mathrm{V_{I}}^+$ & 0.25 & $9.82 \times 10^{2}$\\
        $\mathrm{I_{Cs}}^{2-}$ & 1.10 & $3.50 \times 10^{3}$ & $\mathrm{I_i}^- + \mathrm{V_{Cs}}^{-} $ & $-0.31$ & $1.10 \times 10^{13}$\\
        $\mathrm{Cs_I}^{2+}$ & 1.40 & $4.39 \times 10^{-2}$ & $\mathrm{Cs_{i}}^+ + \mathrm{V_{I}}^+$ & 0.08 & $3.28 \times 10^{7}$\\
        \hline
    \end{tabular}
    
    \textsuperscript{\emph{a}} The specific nonequilibrium conditions are defined by defect formation at an elevated temperature equilibrium at $T=500$K, followed by allowing for recombination through isolation at $T=300$K.
\end{table}

In principle it is possible that a Pb vacancy, $\mathrm{V_{Pb}}^{2-}$, captures two Cs$^+$ ions to form the $\mathrm{[2Cs]_{Pb}}$ antisite, see Figure \ref{fig: structures}(h). Somewhat surprisingly, the most stable charge state at the intrinsic Fermi level of this antisite is $\mathrm{[2Cs]_{Pb}}^+$. Its DFE is, however, $\gtrsim 1$ eV larger than that of the $\mathrm{Cs_{Pb}}^-$ antisite, demonstrating that it is difficult to plant two Cs ions in one Pb lattice position, see Figures \ref{fig: structures}(g) and (h). 

A second possible type of antisite results from placing an anion on a cation position, or vice versa. There are four possibilities, see Figures \ref{fig: structures}(i-l). Again we maintain the nomenclature of antisites, but note that the replacing anion or cation typically does not occupy a lattice site. For instance, in the $\mathrm{I_{Cs}}$ antisite the I ion does not replace the Cs ion at its lattice position, Figure \ref{fig: structures}(k). Instead, it forms a Pb-I-Pb bridge bond nearby, which is a typical bonding configuration for I interstitials \cite{Xue2022perovskites}. In this sense, an antisite is actually a bonding configuration between a vacancy and an interstitial. 

The most stable charge states of these antisites can be guessed from summing the charges of the point defects that can recombine to these antisites. For instance, $\mathrm{Cs_{I}}^{2+}$ and $\mathrm{Pb_{I}}^{3+}$ antisites originate from recombining $\mathrm{Cs_i}^+$, respectively $\mathrm{Pb_i}^{2+}$ interstitials with $\mathrm{V_{I}}^{+}$ vacancies, whereas the $\mathrm{I_{Cs}}^{2-}$ antisite results from recombining an $\mathrm{I_i}^-$ interstitial with a $\mathrm{V_{Cs}}^{-}$ vacancy. $\mathrm{I_{Pb}}^{2-}$ is an exception to this rule; it might be a recombination between an $\mathrm{I_i}^0$ interstitial and a $\mathrm{V_{Pb}}^{2-}$ vacancy. In general, cation-anion antisites lead to unusually high charge states for the defects inserted into the CsPbI$_3$ lattice, Figures \ref{fig: structures} (i-l). This might in part explain their large DFEs, where all cation-anion antisite defects have a DFE that is at least $0.5$ eV larger than that of simple point defects, Figure \ref{fig: DFEs} and Table \ref{table DFEs}. 

In summary, under equilibrium conditions at room temperature, only the formation of cation(Cs)-cation(Pb) antisites is prominent, with $\mathrm{Pb_{Cs}}^+$ presenting a comparable concentration to those of the dominant point defects $\mathrm{Cs_i}^+$ and $\mathrm{V_{Pb}}^{2-}$, and $\mathrm{Cs_{Pb}}^-$ is formed to a lesser extent. Other compound defects, antisites, compound vacancies or interstitials, are not favorable due to their large DFEs.

\subsection{Nonequilibrium Conditions}
Defect concentrations can change drastically under nonequilibrium conditions. Highly nonequilibrium conditions typically occur during the growth of the perovskite crystals. The resulting concentration of defects can then not be simply deduced from the equilibrium relation, Eq. (\ref{eq: defect concentration}). The types and concentrations of defects that occur of course depend on the exact growth conditions. To estimate the potential role played by compound defects, we explore the following model. 

It starts from the assumption that initially defects are created at an elevated temperature, which could reflect an annealing step during the growth process, for instance, with concentrations that can be estimated from Eq. (\ref{eq: defect concentration}). The crystal is then brought to room temperature, where the point defects and compound defects present are allowed to recombine or dissociate, according to Eq. (\ref{eq: mass action}), under the constraints of conservation of the total number of atoms in the defects, Eq. (\ref{eq:conservation}).

\begin{figure}
    \includegraphics[width=14cm]{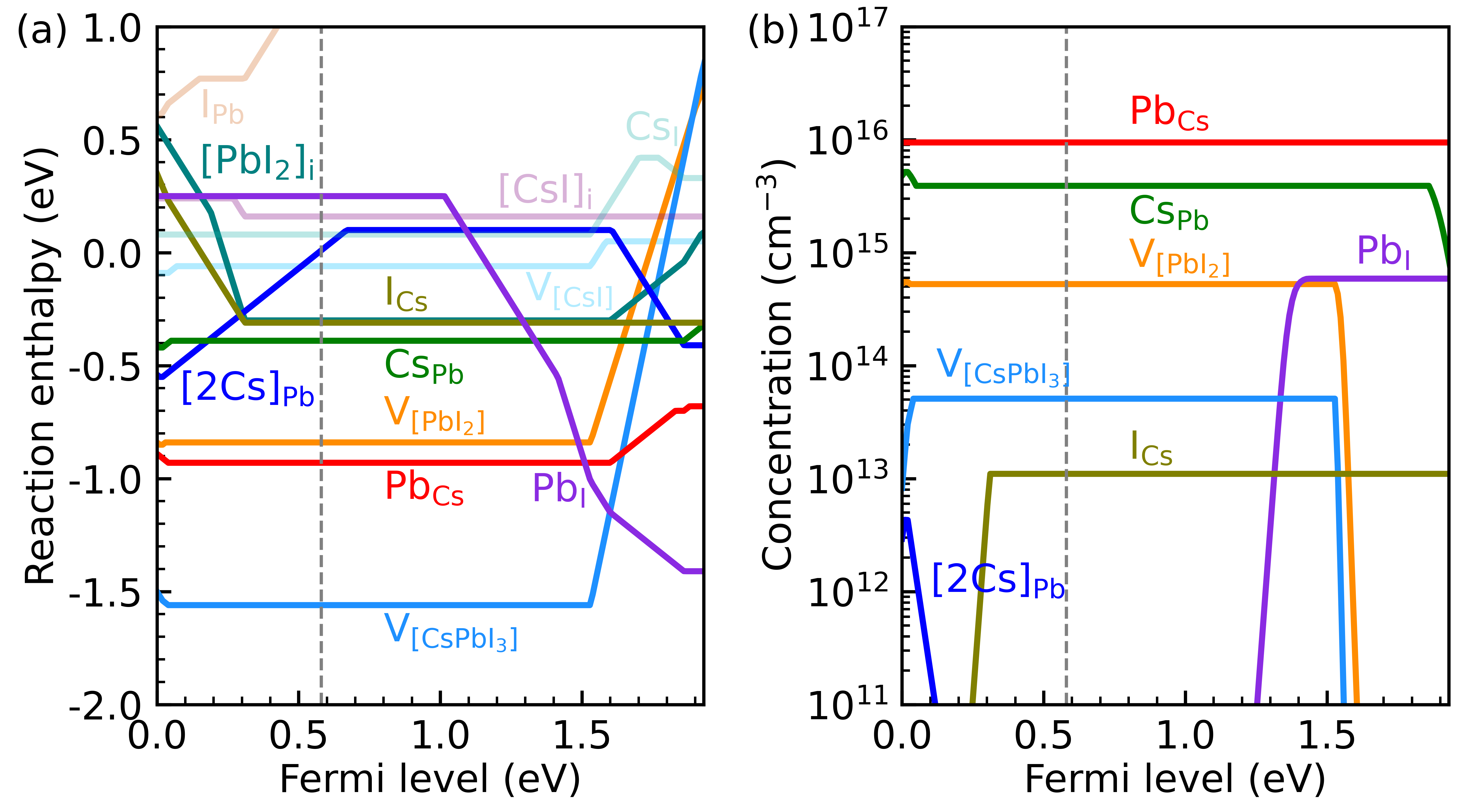}
    \caption{(a) Reaction energies of compound defects, Eq. (\ref{eq: delta H}), as a function of the Fermi level; (b) Concentrations under nonequilibrium conditions, resulting from the law of mass action at room temperature, Eqs. (\ref{eq: mass action}) and (\ref{eq:conservation}), with the initial concentrations of defects determined by equilibrium at $T=500$K. The intrinsic Fermi level is indicated by the vertical dashed gray line.}
    \label{fig: reaction energy}
\end{figure}

A key parameter determining the recombination reaction is the reaction energy, Eq. (\ref{eq: delta H}). Table \ref{table DFEs} shows the reaction energies, calculated at the intrinsic Fermi level, of the recombination reactions that lead to the compound defects, and Figure \ref{fig: reaction energy}(a) shows the reaction energies as a function of the Fermi level. For a recombination reaction to lead to an appreciable concentration of a compound defect, its reaction energy needs to be significantly negative.

Figure \ref{fig: reaction energy}(a) and Table \ref{table DFEs} show that at the intrinsic Fermi level this is the case for the compound vacancies $\mathrm{V_{[CsPbI_3]}}^0$ and $\mathrm{V_{PbI_2}}^0$ and the antisite $\mathrm{Pb_{Cs}}^+$, with reaction energies in the range of $-0.8$ to $-1.5$ eV. The antisites $\mathrm{Cs_{Pb}}^-$ and $\mathrm{I_{Cs}}^{2-}$, as well as the compound interstitial $\mathrm{[PbI_2]_i}^0$ have a moderately negative reaction energy between $-0.3$ and $-0.4$ eV, whereas that of the compound vacancy $\mathrm{V_{CsI}}^0$ is marginally small. The reaction energies of other compound defects, anion-cation antisites (except the mentioned $\mathrm{I_{Cs}}^{2-}$)  and the double antisite $\mathrm{[2Cs]_{Pb}}^+$, or the compound interstitial $\mathrm{[CsI]_i}^0$, are positive, which means that these complexes are not formed in significant concentrations.

However, merely having a negative reaction energy does not imply that a compound defect will form in an appreciable concentration, as formation of a complex necessarily involves a decrease in entropy. Using the law of mass action, Eq. (\ref{eq: mass action}), which is based upon free energies, the effects of entropy are included. At room temperature equilibrium conditions, the most prominent point defects are the Pb vacancy $\mathrm{V_{Pb}}^{2-}$ and the Cs interstitial $\mathrm{Cs_i}^+$, with concentrations of $1.11\times 10^{12}$ cm$^{-3}$, $5.03 \times 10^{11}$ cm$^{-3}$, respectively \cite{Xue2022perovskites}. Under those conditions, all compound defects have a concentration that is at least three orders of magnitude lower, see Figure S1 of the Supporting Information, which means that the loss of entropy involved in their formation reaction essentially prohibits the occurrence of compound defects. The antisite defect $\mathrm{Pb_{Cs}}^+$ is an exception, which forms in a large concentration of $1.71 \times 10^{12}$ cm$^{-3}$ resulting from its low formation energy rather than the recombination of point defects.  

At $T=500$ K the concentrations of the most prominent point defects, $\mathrm{V_{Pb}}^{2-}$ and $\mathrm{Cs_i}^+$, and the cation-cation antisite $\mathrm{Pb_{Cs}}^+$ are raised to $\sim 10^{16}$ cm$^{-3}$, see Figure S2 and Table S2 of the Supporting Information. Based on these initial conditions, the concentrations of compound defects at $T=300$ K (and intrinsic Fermi level,  Eq. (\ref{eq:charge_neutrality})), according to Eqs. (\ref{eq: mass action}) and (\ref{eq:conservation}), are shown in Figure \ref{fig: reaction energy}(b). Most noticeable under these circumstances is that the two cation point defects recombine to form the antisite $\mathrm{Cs_{Pb}}^-$ in a large concentration of $3.88 \times 10^{15}$ cm$^{-3}$. 

A third relatively prominent defect is the compound vacancy $\mathrm{V_{[PbI_2]}}^0$ with a concentration of $5.26 \times 10^{14}$ cm$^{-3}$. The compound vacancy $\mathrm{V_{[CsPbI_3]}}^{0}$ and the anion-cation antisite $\mathrm{I_{Cs}}^{2-}$ occur at lower concentrations of $5.10 \times 10^{13}$ cm$^{-3}$ and $1.10 \times 10^{13}$ cm$^{-3}$, respectively, whereas the concentrations of the other compound defects are much smaller (under intrinsic Fermi level conditions). 

In summary, whereas at equilibrium conditions compound defects are unlikely to form at room temperature, creation of point defects at elevated temperatures and subsequent annealing leads to recombination of point defects, and a prominent appearance of cation-cation antisites $\mathrm{Pb_{Cs}}^+$ and $\mathrm{Cs_{Pb}}^-$. Less important, though still present in appreciable quantities, are the compound vacancies $\mathrm{V_{[PbI_2]}}$ and $\mathrm{V_{[CsPbI_3]}}$, and the anion-cation antisite $\mathrm{I_{Cs}}^{2-}$. 

\subsection{Shifting the Fermi Level}
Nonequilibrium conditions of a different type occur when operating perovskite solar cells. Electrons and holes are produced by light absorption, creating quasi-Fermi levels for electrons and holes that are closer to the band edges than the intrinsic Fermi. The DFEs, Eq. (\ref{eq: DFE}), and therefore the defect concentrations, Eq. (\ref{eq: defect concentration}), are affected by the position of the Fermi level, depending on the charge states of the defects. 

As can be observed in Figures \ref{fig: DFEs}(a,b), the compound vacancies and interstitials maintain their neutral states (and their DFEs) over a large range of Fermi level positions. Only if the Fermi level is close to the conduction band minimum (CBM) does $\mathrm{V_{CsI}}$ become negatively charged, and if the Fermi level is close to the valence band maximum (VBM), $\mathrm{[CsI]_i}$ and $\mathrm{[PbI_2]_i}$ become positively charged. 

The cation-cation antisite $\mathrm{[2Cs]_{Pb}}$, which is positively charged at the intrinsic Fermi level, Figure \ref{fig: DFEs}(c), becomes neutral upon raising the Fermi level, and becomes negatively charged for a Fermi level close to the CBM. The other cation-cation antisites behave similar to simple (charged) point defects, with $\mathrm{Pb_{Cs}}^+$ decreasing its DFE upon lowering the Fermi level, and $\mathrm{Cs_{Pb}}^-$ decreasing its DFE upon raising the Fermi level.

The DFEs of the highly charged cation-anion antisites of course depend strongly on the position of the Fermi level, Figure \ref{fig: DFEs}(d),  the $\mathrm{Cs_{I}}^{2+}$ and $\mathrm{Pb_{I}}^{3+}$ antisites becoming favorable for Fermi level positions close to the VBM, and the $\mathrm{I_{Cs}}^{2-}$ and $\mathrm{I_{Pb}}^{3-}$ becoming more important for Fermi levels close to the CBM.

At the intrinsic Fermi level, or indeed for a Fermi level positioned anywhere in the midgap region, we find that the most stable charge state of a compound defect is simply the sum of the charges of the point defects involved in the recombination reaction, Eq. (\ref{eq: chemical reaction}), 
\begin{equation} \label{eq: chargeconservation}
\alpha_1 q_1 + ... + \alpha_m q_m = \beta q_B
\end{equation}
As long as this holds, the reaction energy does not depend on the exact position of the Fermi level and is constant, see Eqs. (\ref{eq: DFE}) and (\ref{eq: delta H}), which can be observed in Figure \ref{fig: reaction energy}(a) and Figure S3 of the Supporting Information. Consequently, for a Fermi level in this range, the concentrations of the compound defects do not depend upon the exact position of the Fermi level, see Figure \ref{fig: reaction energy}(c).

If the Fermi level is close to the band edges, defects change their charge states, as discussed above. In fact, charge conservation, Eq. (\ref{eq: chargeconservation}) does not necessarily hold, as it becomes energetically more advantageous to accept holes or electrons from the valence or conduction bands by one or more of the defects involved in the reaction. Figure \ref{fig: reaction energy}(a) shows that, as a result of this, reaction energies can change significantly if the Fermi level comes closer to the band edges. As an example, the reaction energy of $\mathrm{[2Cs]_{Pb}}$ decreases if the Fermi level either is close to the VBM or close to the CBM, where this compound defect becomes positively ($\mathrm{[2Cs]_{Pb}}^+$), respectively negatively ($\mathrm{[2Cs]_{Pb}}^{2-}$) charged.

Most remarkable in Figure \ref{fig: reaction energy}(a) is the strong decrease of the reaction energy of the cation-anion antisite $\mathrm{Pb_I}$ if the Fermi level moves upwards from 1.01 eV. At the intrinsic Fermi level, this compound defect is highly charged, $\mathrm{Pb_I}^{3+}$, Figure \ref{fig: DFEs}(d), but upon raising the Fermi level, it becomes energetically advantageous to capture one or more electrons from the conduction band, and lower its reaction energy. Further noticeable is the strong increase of the reaction energies of the compound vacancies $\mathrm{V_{[PbI_2]}}$ and $\mathrm{V_{[CsPbI_3]}}$ for Fermi levels close to the CBM, and for the antisite $\mathrm{I_{Cs}}$ and compound interstitial $\mathrm{[PbI_2]_i}$ for a Fermi level close to the VBM. A detailed description of the reaction and defect formation energies of each compound defect is given in Figure S3 of the Supporting Information.

These changes of the reaction energies upon moving the Fermi level closer to the band edges, have consequences for the defect concentrations, Figure \ref{fig: reaction energy}(b). The cation-cation antisites $\mathrm{Pb_{Cs}}$ and $\mathrm{Cs_{Pb}}$ remain the dominant defect, but for a Fermi level close to the CBM ($E_F>1.6$ eV), the concentration of the compound vacancies $\mathrm{V_{[PbI_2]}}$ and $\mathrm{V_{[CsPbI_3]}}$, which are third and fourth most important defects at midgap Fermi level positions, become negligible. The cation-anion antisite $\mathrm{Pb_I}$ becomes the third most important defect for $E_F>1.3$ eV. For a Fermi level closer to the VBM not much happens, unless $E_F<0.1$ eV, where the antisite $\mathrm{[2Cs]_{Pb}}$ begins to appear in non-negligible concentrations, while the anion-cation antisite $\mathrm{I_{Cs}}$ concentration becomes negligible.

\begin{figure}
    \includegraphics[width=14cm]{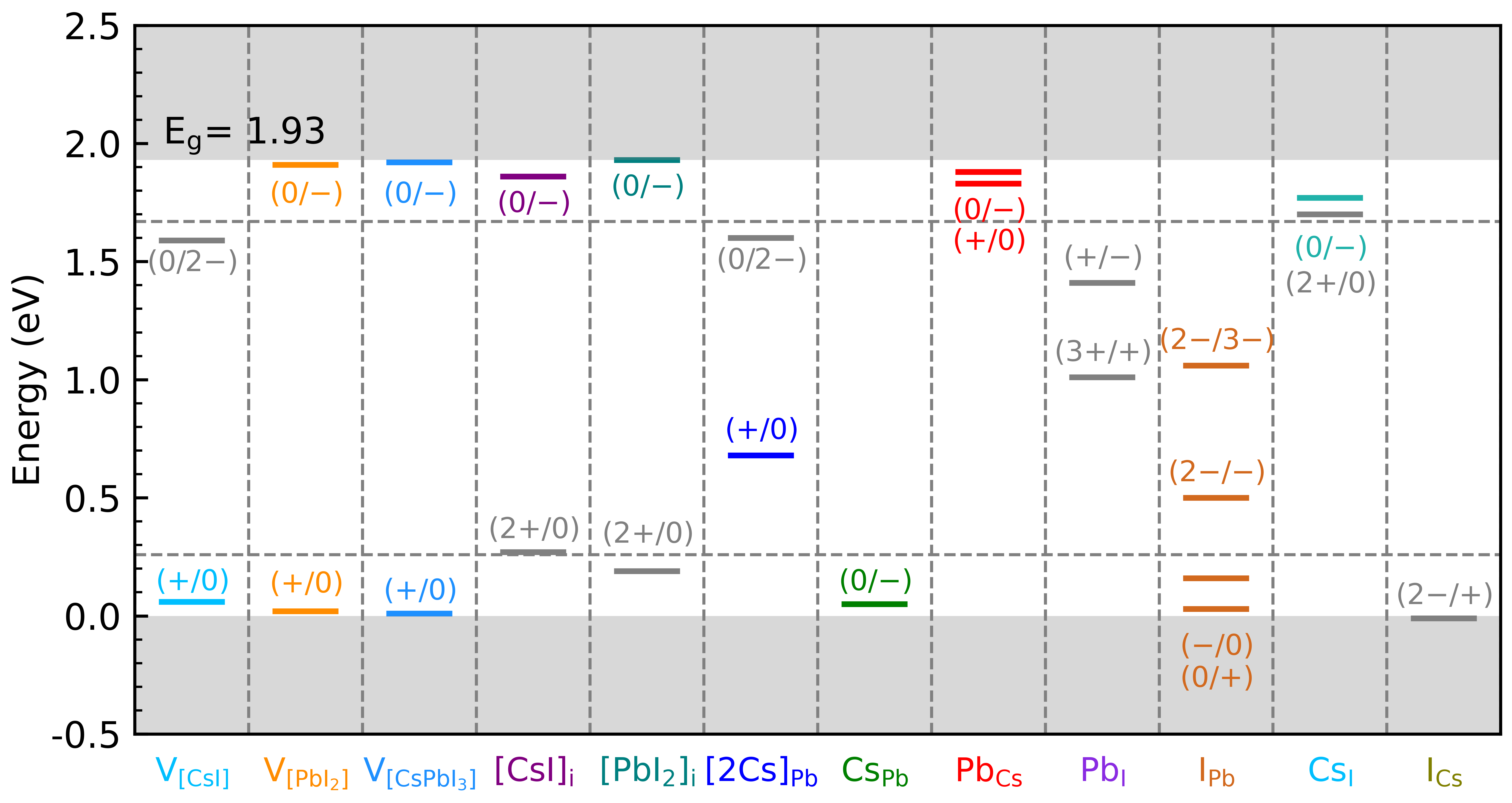}
    \caption{Charge state transition levels of compound defects in CsPbI$_3$. The levels representing a change of a single $\pm e$ are indicated by colored lines. The bottom and top gray areas represent the valence and conduction bands (calculated with the SCAN+rVV10 functional without spin-orbit coupling). The two horizontal dashed lines are 10 $k_\mathrm{B}T$ ($T=300$K) above the VBM and below the CBM, respectively.} \label{fig: CSTLs}
\end{figure}

\subsection{Charge State Transition Levels}
Based on Figure \ref{fig: DFEs} and Eq. (\ref{eq: CSTL}), the CSTLs of compound defects are determined. The results for all defects considered in this paper, are shown in Figure \ref{fig: CSTLs}. The most prominent compound defect, the cation-cation antisite $\mathrm{Pb_{Cs}}$, leads to double shallow donor levels, whereas the antisite $\mathrm{Cs_{Pb}}$ only leads to a shallow acceptor level. The compound vacancies $\mathrm{V_{[PbI_2]}}$ and $\mathrm{V_{[CsPbI_3]}}$ have both a shallow donor, as well as a shallow acceptor level. The anion-cation antisite $\mathrm{I_{Cs}}$ has no levels inside the band gap. 

At the intrinsic Fermi level, or indeed if the Fermi level is well inside the band gap, these are all compound defects that can occur in appreciable quantities, see Figure \ref{fig: reaction energy}(b). If the Fermi level is close to the CBM, the concentration of $\mathrm{Pb_{I}}$ antisites increases. Although this antisite introduces two deep levels inside the band gap, both of these levels involve a change in charge state of two electrons, i.e., $3+/1+$ and $+/-$. These levels are likely to be much less active than donor or acceptor levels associated with a change of one in charge state, as the probability of trapping two electrons simultaneously is very low \cite{Tress2017, Meggiolaro2018review}. If the Fermi level becomes extremely close to the VBM, the concentration of $\mathrm{[2Cs]_{Pb}}$ becomes non-negligible. As its CSTL (+/0) is well inside the band gap, this compound defect forms a deep trap, which can act as a harmful recombination center. 

Besides the defects discussed in the previous two paragraphs, all other defects occur in such negligible quantities, that their electronic impact is negligible. In fact, the only compound defect considered in this paper that forms a series of deep trap levels, which is the anion-cation antisite $\mathrm{I_{Pb}}$, Figure \ref{fig: reaction energy}(b), has a very large positive reaction energy, Table \ref{table DFEs}, so it does not form under practical conditions.

In summary, the relatively abundant compound defects either form shallow donor or acceptor levels ($\mathrm{Pb_{Cs}}$, $\mathrm{Cs_{Pb}}$, $\mathrm{V_{[PbI_2]}}$, $\mathrm{V_{[CsPbI_3]}}$, $\mathrm{I_{Cs}}$,) or electronically not very active levels ($\mathrm{Pb_{I}}$). Only under relatively extreme conditions, with a Fermi level very close to the VBM, the compound defect $\mathrm{[2Cs]_{Pb}}$ can form, which has a deep trap level.

\section{Conclusions}

To conclude, we have studied the formation of compound defects in the archetype inorganic halide perovskite CsPbI$_3$ by means of DFT calculations using the accurate and efficient SCAN+rVV10 functional. Considering compound vacancies, $\mathrm{V_{[CsI]}}$, $\mathrm{V_{[PbI_2]}}$, and $\mathrm{V_{[CsPbI_3]}}$, compound interstitials $\mathrm{[CsI]_i}$ and $\mathrm{[PbI_2]_i}$, cation-cation antisites, $\mathrm{Pb_{Cs}}$, $\mathrm{Cs_{Pb}}$, $\mathrm{[2Cs]_{Pb}}$, and anion-cation antisites $\mathrm{I_{Cs}}$, $\mathrm{I_{Pb}}$, $\mathrm{Cs_I}$, $\mathrm{Pb_I}$, we evaluate their formation under equilibrium conditions, and under conditions that reflect their formation as recombination reactions of simple point defects.    

Although the energies of several of these recombination reactions are favorable, under equilibrium conditions at room temperature, only the formation of the antisite where Pb substitutes Cs is prominent, and the concentrations of point defects are too small to give any appreciable amount of other compound defects. However, under nonequilibrium conditions, mimicked by a high temperature annealing step, several types of compound defects can be formed in significant concentrations. Most prominent are the cation-cation antisites $\mathrm{Pb_{Cs}}^+$ and $\mathrm{Cs_{Pb}}^-$, with concentrations comparable to those of the dominant point defects  $\mathrm{Cs_{i}}^+$ and $\mathrm{V_{Pb}}^{2-}$. Smaller amounts of the compound vacancies $\mathrm{V_{[PbI_2]}}^0$ and $\mathrm{V_{[CsPbI_3]}}^0$, and the anion-cation antisite $\mathrm{I_{Cs}}^{2-}$ can be observed, whereas the concentrations of other defects are negligible.

Under solar cell operating conditions the (quasi) Fermi level can shift to the proximity of the VBM and CBM, which promotes the formation of certain compound defects, and suppresses that of others. If the Fermi level is close to the CBM, the formation of $\mathrm{V_{[PbI_2]}}^0$ and $\mathrm{V_{[CsPbI_3]}}^0$ is suppressed, and that of the cation-anion antisite $\mathrm{Pb_{I}}^{-}$ is promoted, whereas if the Fermi level is close to the VBM, the formation of $\mathrm{I_{Cs}}^{2-}$ is suppressed, and that of $\mathrm{[2Cs]_{Pb}}^+$ is promoted. The other defects are less affected by a change in Fermi level.

The antisites and compound vacancies that can occur in appreciable concentrations ($\mathrm{Pb_{Cs}}$, $\mathrm{Cs_{Pb}}$, $\mathrm{I_{Cs}}$, $\mathrm{V_{[PbI_2]}}$ and $\mathrm{V_{[CsPbI_3]}}$) create shallow trap levels only. The antisite $\mathrm{Pb_{I}}$ creates several deep levels, which are, however, not very active electronically, as their charge state transition involves the arrival of two electrons simultaneously. Only the compound defect $\mathrm{[2Cs]_{Pb}}$ leads to a deep trap level. However, as discussed above, this defect is only likely to form if the Fermi level is very close to the VBM. These results illustrate the exemplary electronic tolerance of halide perovskites towards the presence of defects.

\begin{acknowledgement}

H.X. acknowledges funding from the China Scholarship Council (CSC, No. 201806420038). S.T. acknowledges funding from the Computational Sciences for Energy Research (CSER) tenure track program of Shell and NWO (Project No. 15CST04-2) and the NWO START-UP grant from the Netherlands.

\end{acknowledgement}

\begin{suppinfo}

A list of number of possible defect sites and counting rule for each compound defect; Concentration of compound defects at 300 K; Temperature dependence of concentrations of point defects; Table of concentrations of point defects at 300 K and 500 K; Detailed comparison of reaction energies and formation energies of each compound defect.

\end{suppinfo}

\bibliography{Manuscript}

\end{document}



\clearpage

\tableofcontents

\clearpage

\section{Density of possible defect sites}

In the main text, $c_0(D^q)$ in Equations (6) and (7) defines the density of possible sites for the defect, including orientational degrees of freedom. It is calculated from 
\begin{equation}
    c_0(D^q) = \frac{n(D^q)}{V_{f.u.}}, 
    \label{eq: density of possible defect sites}
\end{equation}
where $n(D^q)$ is the number of possible sites and orientations for the defect $D^q$ per formula unit of CsPbI$_3$, and $V_{f.u.}$ is the volume per formula unit, which is calculated to be 237 \AA$^3$ or $2.37 \times 10^{-22}$ cm$^3$. The values of $n(D^q)$ for each compound defect are given in Table \ref{table: number of possible sites for defect}.

\begin{longtblr}[
  caption = {Number of possible sites for each type of compound defect and the counting rule used.},
  label = {table: number of possible sites for defect},
]{
  colspec = {X[0.5,l]X[0.5,l]X[4,l]},
  rowhead = 1,
} 
    \hline
        Defect & $n(D^q)$ & Remarks \\
        \hline
        \multicolumn{3}{l}{\underline{Vacancies}} \\
        $\mathrm{V_{CsI}}$ & 3 & 1 Cs in the center of of the CsPbI$_3$ cube, with 12 possibilities for iodine anions in the middle of each edge to be an iodine vacancy, and each edge is shared by 4 cubes; so $1 \times 12/4 = 3$.\\
        $\mathrm{V_{PbI_2}}$ & 1 & 8 Pb cations at the corners of the cube, with each corner shared by 8 cubes; 2 iodine ions with one at the a-site and another at the e-site, and each of the two possibilities of combination is shared by two faces; so $8/8 \times 2/2 = 1$.\\
        $\mathrm{V_{CsPbI_3}}$ & 1 & 8 Pb cations at the corners of a cube, with three iodine ions forming a corner of a cube; so $8/8 \times 1 = 1$.\\
        \multicolumn{3}{l}{\underline{Interstitials}} \\
        $\mathrm{[CsI]_i}$ & 3 & The Cs interstitial occupies a face of a CsPbI$_3$ cube, with 6 faces per cube, and each face shared by 2 cubes; the Cs interstitial is accompanied by an iodine interstitial occupies a site next to a lattice iodine anion in the middle of one of the four edges of a face, and each edge is shared by 4 cube; so $ 6/2 \times 4/4 = 3 $.\\
        $\mathrm{[PbI_2]_i}$ & 6 & The Pb interstitial occupies a face of a CsPbI$_3$ cube, with 6 faces per cube, and each face shared by 2 cubes; the Pb interstitial is accompanied by two iodine interstitials, with the I-Pb-I plane parallel to the in-plane or out-of-plane; so $ 6/2 \times 2 = 6 $.\\
        \multicolumn{3}{l}{\underline{Antisites (cation-cation)}} \\
        $\mathrm{Pb_{Cs}}$ & 1 & 1 Pb cation replaces the Cs cation in the center of a cube. \\
        $\mathrm{Cs_{Pb}}$ & 1 & 1 Cs cation replaces one of the 8 Pb cations at the corners of the cube, with each corner shared by 8 cubes; so $ 8/8 = 1 $.\\
        $\mathrm{[2Cs]_{Pb}}$ & 1 & Similar to the $\mathrm{Cs_{Pb}}$.\\
        \multicolumn{3}{l}{\underline{Antisites (cation-anion)}} \\
        $\mathrm{I_{Pb}}$ & 1 & Similar to the $\mathrm{Cs_{Pb}}$.\\
        $\mathrm{Pb_{I}}$ & 3 & the Pb interstitial has a square pyramidal bonding to 5 surrounding I anions; the iodine anion in the middle of the edge opposite to the one where the iodine vacancy is serves as an apex of a pyramid, and all pyramids are inside the cube; so $ 12/4 = 3$. \\
        $\mathrm{I_{Cs}}$ & 3 & A Cs vacancy, with an iodine interstitial occupies a site next to a lattice iodine anion in the middle of each edge, and each edge is shared by 4 cube; so $ 1 \times 12/4 = 3 $.\\
        $\mathrm{Cs_I}$ & 3 & The Cs interstitial occupies a face of a CsPbI$_3$ cube, with 6 faces per cube, and each face shared by 2 cubes; the iodine vacancy can be one of the four lattice iodine ions in the middle of each edge surrounding the Cs interstitial, with each edge shared by 4 cubes; so $ 6/2 \times 4/4 = 3 $. \\
    \hline
\end{longtblr}

\clearpage

\section{Concentrations of compound defects at 300 K}
\begin{figure} [h]
    \centering
    \includegraphics[width=0.55\textwidth]{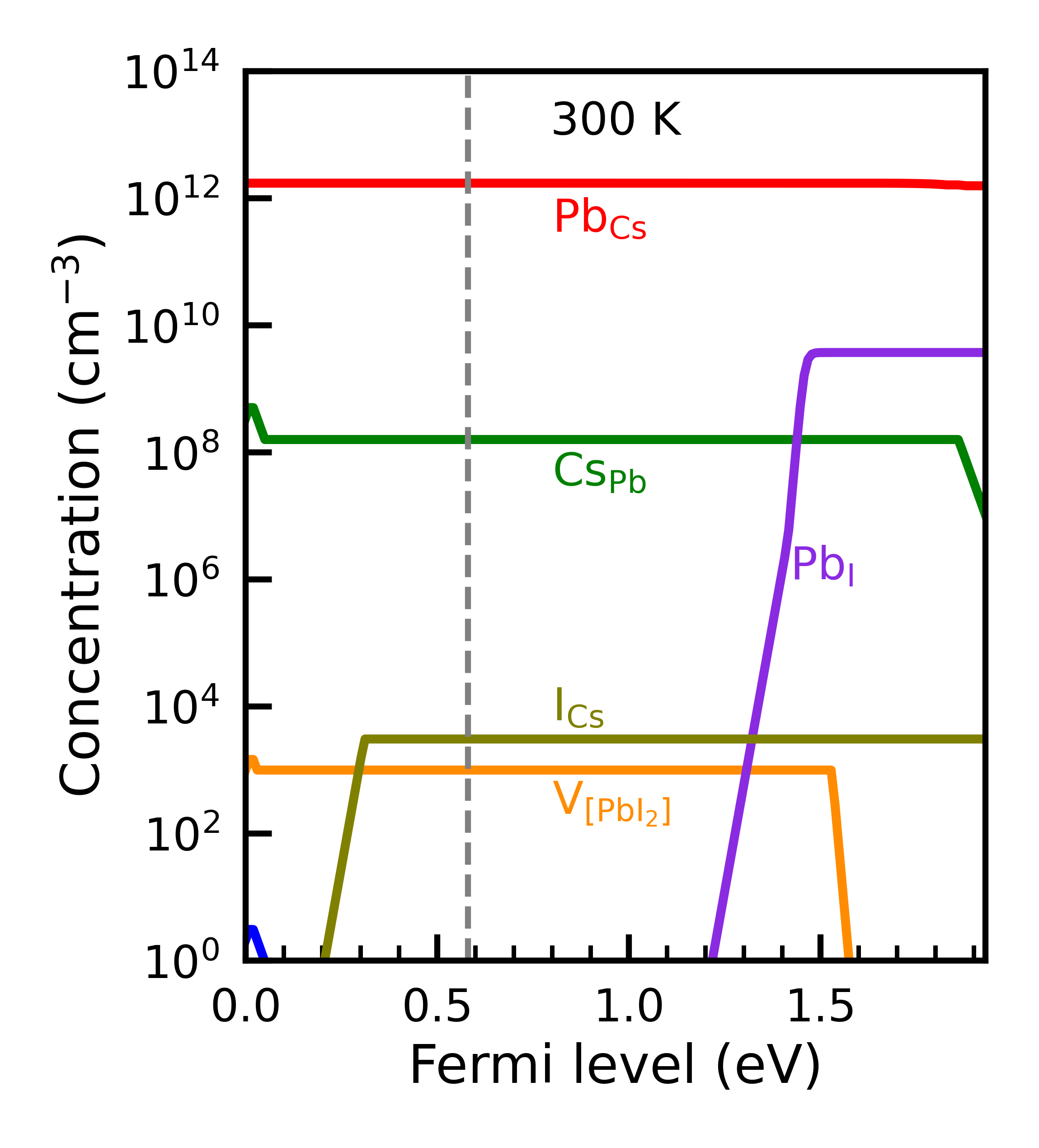}
    \caption{Concentrations resulting from the law of mass action at room temperature, with the initial concentrations of defects determined at equilibrium conditions at $T=300$ K.}
    \label{fig: 300K}
\end{figure}

\clearpage

\section{Temperature dependence of concentrations of point defects}
\begin{figure} [h]
    \centering
    \includegraphics[width=0.55\textwidth]{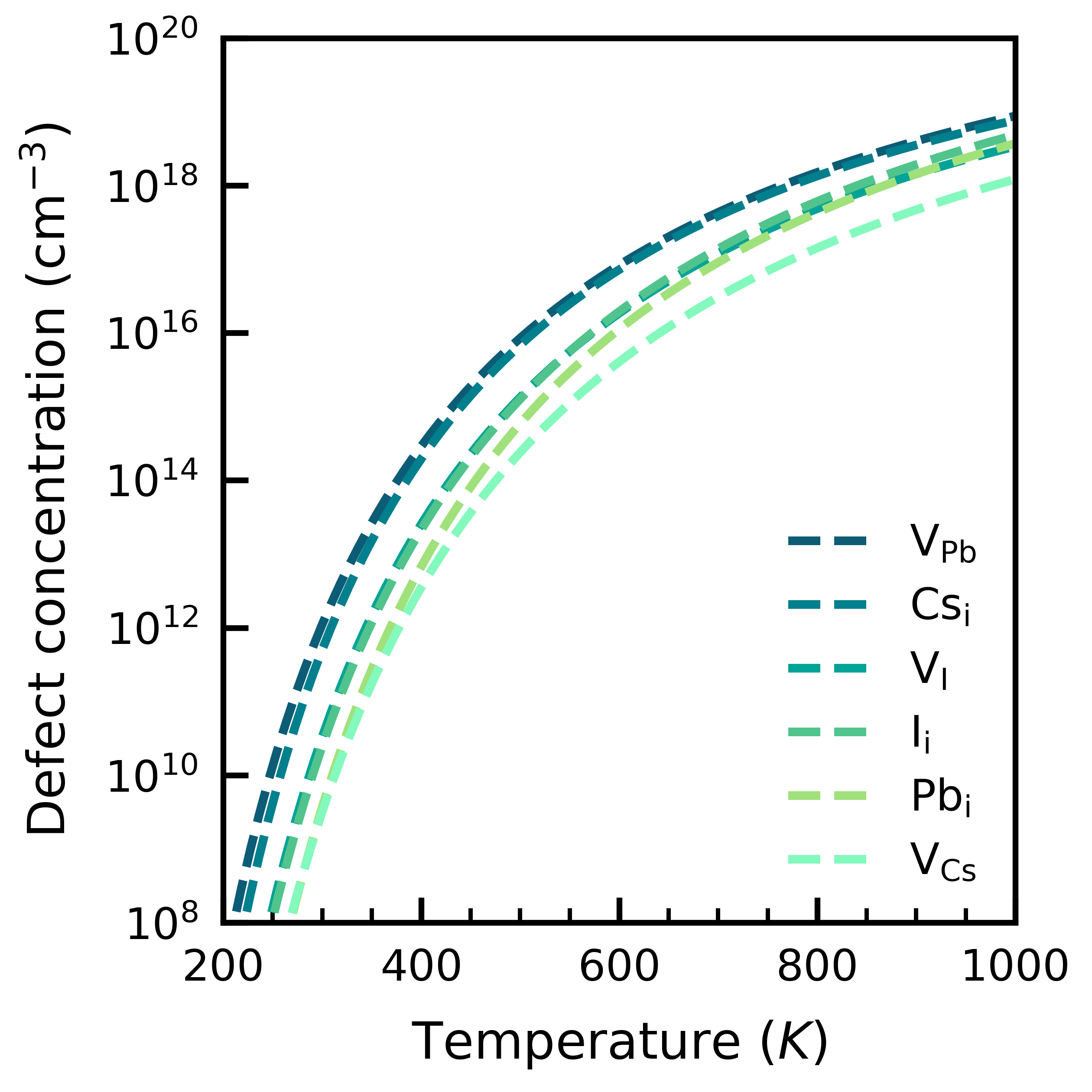}
    \caption{Equilibrium concentrations of point defects as a function of temperature.}
    \label{fig: concentration and temperature}
\end{figure}

\begin{table}[h!]
    \caption{Equilibrium concentration of point defects calculated at 300 K and 500 K.}
    \label{tab: concentraion of point defects}
    \centering
    \begin{tabular}{lcc}
        \hline
        \multirow{2}{*}{Defect} & \multicolumn{2}{c}{Concentration (cm$^{-3}$)} \\
        \cline{2-3}
         & 300 K & 500 K \\
         \hline
        $\mathrm{V_{Cs}}^-$ & $3.21\times10^{9}$ & $2.40\times10^{14}$ \\
        $\mathrm{V_{Pb}}^{2-}$ & $1.11\times10^{12}$ & $8.53\times10^{15}$ \\
        $\mathrm{V_{I}}^+$ & $3.31\times10^{10}$ & $1.34\times10^{15}$ \\
        $\mathrm{Cs_i}^+$ & $5.03\times10^{11}$ & $6.85\times10^{15}$ \\
        $\mathrm{Pb_i}^{2+}$ & $3.70\times10^{9}$ & $5.89\times10^{14}$ \\
        $\mathrm{I_i}^-$ & $2.48\times10^{10}$ & $1.27\times10^{15}$ \\
        \hline
    \end{tabular}
\end{table}

\clearpage

\section{Reaction energies and defect formation energies}

\begin{figure} [h]
    \includegraphics[width=1\columnwidth]{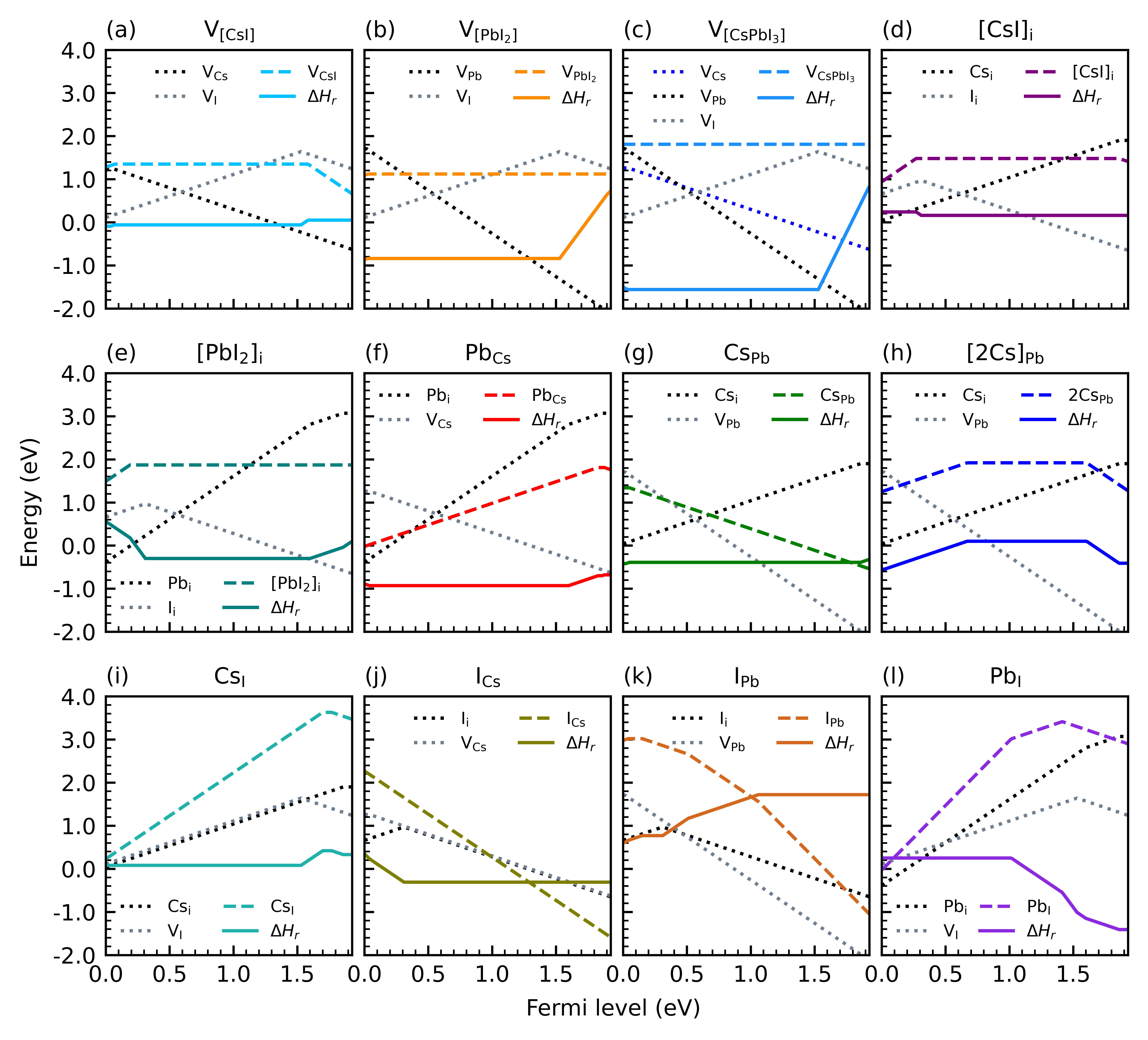}
    \caption{Reaction energies ($\Delta H_r$) and defect formation energies (DFEs) as a function of Fermi level of compound defects. DFEs of the point defects are represented by dotted lines, while those of the compound defects are represented by dashed lines. The reaction energies are represented by solid lines. }
    \label{fig: reaction enthalpy and DFE}
\end{figure}
